\documentclass[lettersize,journal]{IEEEtran}
\usepackage{amsmath,amssymb,amsfonts,amsthm, amssymb}
\usepackage{mathtools,array,stfloats,graphicx,url,verbatim,cite,textcomp,tabularx,mathrsfs}
\usepackage[table]{xcolor} 
\usepackage{colortbl}
\usepackage{algorithmic}
\usepackage{subcaption}
\usepackage{hyperref}
\usepackage[ruled]{algorithm}

\theoremstyle{remark}

\newcommand\norm[1]{\left\lVert#1\right\rVert}

\begin{document}
\title{Distributed Linear Bandits for Freshness-Aware Preprocessing in MEC-Enabled Digital Twin Networks\\
\thanks{This work was supported in part by the Federal Ministry of Research,
Technology and Space (BMFTR) under Grant 16KIS2411 and in part by German
Research Foundation (DFG) under Grant MA7111/6-1

Mariam Yahya and Setareh Maghsudi are with the Faculty of Electrical Engineering and Information Technology, Ruhr-University Bochum, 44801 Bochum, Germany (e-mails: \{mariam.yahya, setareh.maghsudi\}@ruhr-uni-bochum.de).
}
}
\author{  Mariam Yahya and Setareh Maghsudi\\
\IEEEauthorblockA{
    Faculty of Electrical Engineering and Information Technology, Ruhr-University Bochum, Germany}
}
\maketitle
\begin{abstract}
Digital twins (DTs) require timely updates from Internet of Things (IoT) devices to accurately represent the state of the corresponding physical systems, but transmitting large data volumes can compromise freshness. Moreover, different task classes have distinct computational demands and freshness requirements, while stochastic task generation and sizes complicate resource allocation. In this paper, we propose a freshness-aware framework in which multi-access edge computing (MEC) servers preprocess IoT data, reducing transmission volume and DT-server processing load. The framework establishes mutual support between DT and MEC servers. The DTs determine feasible CPU-allocation strategies guaranteeing class-specific age-of-information (AoI) requirements under specified workload, delay, and update-generation bounds. MEC servers collaboratively identify the feasible strategy minimizing expected freshness-normalized preprocessing delay. We formulate CPU-strategy selection as a fixed-confidence distributed best-arm identification problem in linear bandits, with MEC servers as agents and CPU allocations across task classes as arms.  We employ DistLinGapE, a distributed collaborative learning method, to coordinate exploration and share observations. Numerical results show that the method identifies the optimal feasible strategy and reduces preprocessing delay when deployed under stationary task-class assignments and workload statistics. Collaboration yields an identification speedup nearly proportional to the number of participating MEC servers in the evaluated settings.
\end{abstract}
\begin{IEEEkeywords}
    Multi-access edge computing (MEC), Age-of-Information (AoI), digital twins (DT), linear bandits, distributed learning.
\end{IEEEkeywords}
\section{Introduction}
The rapid growth of Internet of Things (IoT) networks has accelerated the use of digital twins (DTs), which are virtual representations of physical systems, in computation-intensive applications such as vehicular networks \cite{chai2026cost}, industrial automation \cite{bozkaya2025digital}, and smart healthcare \cite{kabir2025digital}. Through continuous synchronization with physical systems, DTs can support system monitoring, optimization, and autonomous operation \cite{chai2026cost, bozkaya2025digital,kabir2025digital,alkhateeb2023real, li2025delay,noroozi2025age}. Maintaining an accurate DT, however, requires sensing devices to provide timely data that reflects the current state of the physical system.

A common approach updates the virtual DT representations and uses them to guide operational decisions in the physical system, such as resource allocation in edge networks \cite{bozkaya2025digital,zhou2025digital}. However, updating the DT itself requires communication and computational resources. Transmitting large volumes of sensing data to the DT server can cause substantial communication overhead and processing delays \cite{li2025delay,sanjalawe2026bridging}. The resulting synchronization delays may cause the DT to rely on outdated information, reducing its usefulness in real-time applications. Moreover, depending on the application, updating the DT may require only a subset of the sensing data or information extracted from it \cite{krantzik2025optimizing,tang2024uav,li2025delay,sanjalawe2026bridging}.

Preprocessing sensing data before updating the DT can help address this challenge. Techniques such as data compression \cite{krantzik2025optimizing}, feature extraction \cite{tang2024uav}, and semantic communication \cite{li2025delay,sanjalawe2026bridging} can reduce the volume of forwarded data and, depending on the application, the processing load at the DT server. These benefits motivate the use of multi-access edge computing (MEC) servers to preprocess data closer to IoT devices.

IoT devices or clusters can transmit sensing tasks from multiple classes to MEC servers for preprocessing. These classes may differ in their preprocessing needs, computational demands, and freshness requirements \cite{tzeng2025age, park2020centralized}. Age of information (AoI) provides a measure of the freshness of the information represented by a DT and can be used to specify class-specific freshness requirements.

Allocating computational resources for preprocessing is challenging when task generation and task sizes are uncertain \cite{zhou2025digital,choi2019latency}, as the allocation must accommodate different computational demands while keeping DTs sufficiently up to date. Uniform or random CPU allocation across task classes may be inefficient because it does not account for differences in workloads and freshness requirements. This motivates learning approaches that use observed preprocessing delays to identify effective CPU allocations.

\subsection{Paper Contributions}
We propose a framework that establishes mutual support between the DT and the MEC servers. The MEC servers preprocess sensing data before forwarding the results to the DT server, while the DT guides their CPU-allocation decisions through a two-stage process. First, the DT determines a feasible set of CPU-allocation strategies that guaranty class-specific AoI requirements under specified workload, delay, and update-generation bounds. Second, the MEC servers collaboratively identify the strategy within this set that minimizes the expected freshness-normalized preprocessing delay.

We formulate CPU-allocation strategy selection as a fixed-confidence distributed best-arm identification (BAI) problem in linear bandits. MEC servers solving a shared allocation problem combine their observations to identify the best strategy with a predefined confidence using as few observations as possible. The linear structure allows observations from one strategy to inform the evaluation of others. We employ the distributed linear gap-based exploration (DistLinGapE) algorithm \cite{yahya2026service} to perform this collaborative identification.

The main contributions of this work are summarized as follows:
\begin{itemize}
    \item We develop a freshness-aware MEC preprocessing framework in which the DTs determine feasible CPU-allocation strategies that satisfy class-specific AoI requirements.
    
    \item We formulate CPU-allocation strategy selection as a fixed-confidence distributed linear BAI problem, aiming to identify the feasible strategy that minimizes the expected freshness-normalized preprocessing delay under unknown mean workloads and task generation rates.
    
    \item We employ DistLinGapE to enable MEC servers to collaboratively identify the best feasible CPU allocation.
    
    \item The numerical results show that the DistLinGapE is effective in finding the best CPU allocation strategy with a predetermined confidence. Additionally, we show that collaboration among the MEC-servers speedup the learning process by reducing the number of learning rounds required per server. 
\end{itemize}

\subsection{Paper Organization}
This paper is organized as follows. Section~\ref{sec:related_work} reviews the related work on DTs, AoI, and bandit learning. Section~\ref{sec:system_model} introduces the system model, and Section~\ref{sec:DT_fresh} presents the DT freshness requirements and strategy-update mechanism. Section~\ref{sec:problem_formulation} formulates the optimization problem, while Section~\ref{sec:structural_model} develops its linear-bandit BAI representation. Section~\ref{sec:collaborative_learning} presents the solution approach using DistLinGapE. Numerical results are presented in Section~\ref{sec:numerical_results}, and Section~\ref{sec:conclusion} concludes the paper.

\section{Background and Related Work}
\label{sec:related_work}

\subsection{Digital Twins and Age of Information}

Research on DT-enabled networks considers two different but complementary directions: the first uses DTs to support physical-system monitoring, optimization, or operation, while the second focuses on maintaining timely DT representations through methods such as  resource allocation, DT placement and migration, and update scheduling. These directions are closely connected, as the effectiveness of DT-assisted decisions depends on the freshness of the information available to the DT. Our work connects them through mutual support between the DT and MEC servers.

Our work considers fixed DT placement and focuses on allocating computational resources to reduce preprocessing delay while satisfying freshness requirements. Another research direction investigates DT placement and migration to determine where DTs should be hosted and when their locations should change to improve synchronization \cite{li2023digital, noroozi2025age, Guo2024}. Dynamic placement adjusts DT locations to changing network conditions  to improve synchronization efficiency \cite{li2023digital}. For mobile physical systems, Noroozi et al. \cite{noroozi2025age} formulate DT migration as a Markovian stopping problem and use dynamic programming to reduce the AoI observed when an application requests data. Considering both freshness and energy consumption, Guo et al. \cite{Guo2024} jointly optimize edge association, transmission power, and DT deployment through a lightweight online algorithm. 

Collaboration in DT-enabled networks has been studied both for providing shared services and for coordinating resource allocation. Although MEC servers and DTs may operate independently, applications involving multiple physical objects can benefit from exchanging information and coordination. For example, Li et al. \cite{li2024aoi} consider DT network slicing, where worker DTs process data from physical objects and a master DT aggregates their outputs to support a shared service. They develop approximation and online algorithms to maximize service utility based on AoI-dependent model quality, subject to delay and resource constraints. Collaboration can also support privacy preservation. Heydari et al. \cite{heydari2026channel} propose a federated optimization framework in which a scheduler coordinates with DTs to minimize transmission time or maximize the amount of uploaded data under energy constraints, without revealing the DTs' private information. In our framework, the DTs of the IoT clusters reside on a common DT server, which may be a MEC server or a separate server. This enables coordinated specification of class-specific freshness thresholds and workload and delay bounds. Collaboration takes place among the MEC servers, which share observations to accelerate the identification of an effective preprocessing CPU allocation.

A major role of the DT is to assist in resource allocation by using virtual representations to make decisions in the physical network. In our work, the DT does not directly select the resource allocation. Instead, it determines a set of feasible allocations based on task freshness requirements and specified workload and delay bounds. The MEC servers then collaboratively identify the best allocation within this set. Other studies use DTs to optimize scheduling and resource-allocation decisions directly. Bozkaya et al. \cite{bozkaya2025digital} propose a genetic algorithm-based scheduler to minimize AoI and task completion time across industrial IoT devices and MEC servers. Huynh et al. \cite{van2023distributed} jointly optimize bandwidth, transmission power, task-offloading fractions, and the processing rates of user devices and edge servers. They develop a centralized solution based on successive convex approximation, followed by a distributed solution.

Beyond resource allocation, some studies also optimize how information is processed for DT synchronization \cite{krantzik2025optimizing,tang2024uav,li2025delay,sanjalawe2026bridging}. For example, Li et al. \cite{li2025delay} jointly optimize semantic extraction, transmission power, and computational resources using a soft actor-critic algorithm to minimize average synchronization latency under deadline, energy, and resource constraints. Other studies use data compression \cite{krantzik2025optimizing} or semantic feature extraction \cite{tang2024uav}. Our work considers MEC-side preprocessing for data-volume reduction without restricting the framework to a specific application or preprocessing technique. For given preprocessing operations, we allocate CPU resources across task classes with heterogeneous computational demands and freshness requirements.

Uncertainty in operating conditions motivates learning-based resource allocation when accurate system or workload information is unavailable \cite{wirth2026risk,bhardwaj2023cilantro,wang2024autothrottle,ai2025joint}. Wirth et al. \cite{wirth2026risk} use a risk-aware bandit algorithm for radio-access and DT-placement selection under changing conditions. Ai et al. \cite{ai2025joint} use a multi-armed bandit strategy to allocate computing resources between model retraining and inference services in DT-assisted MEC. In our setting, the  preprocessing workloads are stochastic with unknown distributions. MEC servers therefore learn allocation performance from observed delays and share observations to accelerate identification within a freshness-feasible set. This motivates the bandit learning approach discussed next.

\subsection{Bandit Learning and Best-Arm Identification}
Bandit learning is a sequential decision-making framework in which a learner, called an agent, selects actions, known as arms, with unknown reward distributions and updates its beliefs from observed rewards to guide subsequent selections. One objective is to minimize cumulative regret, which is the difference between the expected cumulative reward from consistently selecting an optimal arm and that from the selected arms \cite{lattimore2020bandit,abbasi2011improved}. Alternatively, BAI focuses on recommending the best arm after an exploration phase \cite{audibert2010best,xu2018fully}. In the fixed-confidence setting, the objective is to identify an optimal arm with a predefined confidence, using as few observations as possible. The latter approach suits our problem, where CPU allocations are explored until the best feasible allocation is identified with the required confidence and then deployed until the network configuration changes.

Linear bandits have a shared structure across arms, it expresses each arm's expected reward as the inner product of a known feature vector and an unknown parameter vector. This structure allows observations from one arm to inform estimates of other arms, potentially reducing the exploration required for identification. Xu et al. \cite{xu2018fully} propose the linear gap-based exploration (LinGapE) algorithm for fixed-confidence linear BAI. The algorithm is adaptive, meaning that it uses reward observations and confidence bounds to select the next best arm  to eventually identify the best arm in the least number of rounds, a more efficient algorithm compared to static approaches \cite{soare2014best}. The DistLinGapE \cite{yahya2026service} extends this approach to a distributed setting in which multiple agents collaborate by sharing learning updates. Each agent updates its model estimates using local data, and then exchanges the updates with other agents in each global round creating a tradeoff between communication and local computation.

In our problem, uncertainty in task generation and task sizes makes preprocessing delays unknown. We model the CPU-strategy selection as a fixed-confidence linear BAI problem, where each CPU allocation in the set of freshness-feasible allocations determined by the DTs represents an arm. Each MEC server acts as an agent that selects and deploys an allocation and observes the resulting delay. We employ DistLinGapE to coordinate learning among MEC servers. Servers observe local delays under the selected strategies and exchange learning updates. Under suitable conditions, sharing observations reduces learning rounds  per agent compared with independent learning.

\section{System Model}
\label{sec:system_model}
We first introduce the notation used in this work.\\
\textbf{Notation:} We use boldface lowercase and uppercase letters to represent vectors and matrices, respectively. For a positive integer $K$, $[K]$ denotes the set $\{1,\dots,K\}$. Furthermore, $\norm{\boldsymbol{x}}_p$ denotes the $p$-norm of a vector $\boldsymbol{x}$. For a symmetric positive-definite matrix $\boldsymbol{A}$, the weighted $2$-norm is defined as $\norm{\boldsymbol{x}}_{\boldsymbol{A}}=\sqrt{\boldsymbol{x}^{\top}\boldsymbol{A}\boldsymbol{x}}$. We use $\mathbb{P}[\cdot]$ to denote the probability of the event inside the brackets.

\begin{figure}[h]
    \centering
    \includegraphics[width=0.9\linewidth]{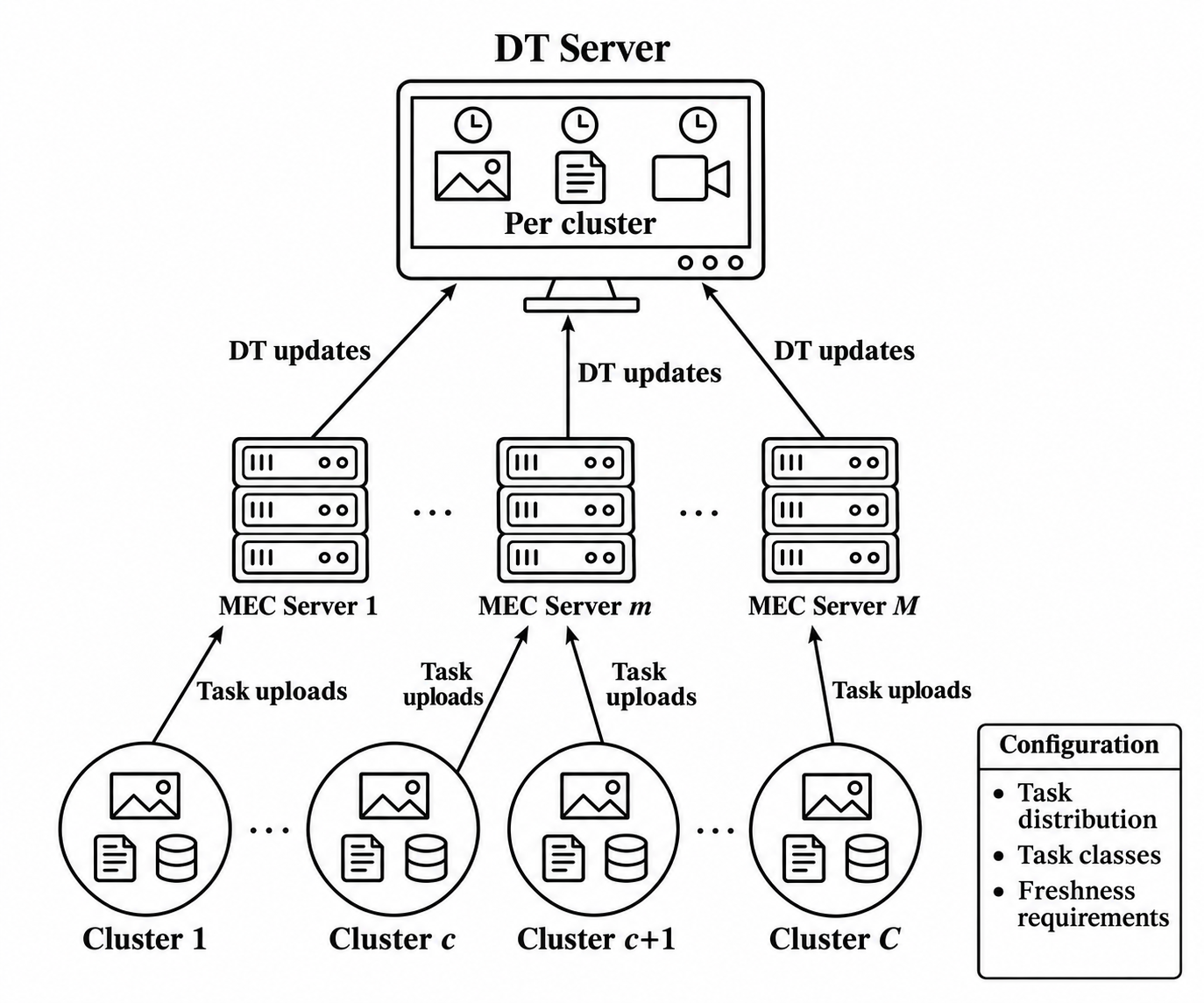}
    \caption{The system model.}
    \label{fig:system_model}
\end{figure}

We consider an IoT network consisting of a set $\mathcal C$ of $C$ clusters, indexed by $c\in[C]$, and a set $\mathcal M$ of $M$ MEC servers, indexed by $m\in[M]$. Each IoT cluster has a corresponding DT that virtually represents the physical system and is updated as new information becomes available. The DTs of all clusters are maintained at a common server, referred to as the DT server, which may be one of the MEC servers or a separate server with resources dedicated to DT operation. The system model is illustrated in Fig.~\ref{fig:system_model}.

\subsection{Cluster Association and Homogeneous MEC Servers}

Each cluster is associated with one MEC server, and we assume that this association is based on proximity. Let $\mathcal C_m$ denote the set of clusters associated with server $m\in[M]$. We consider applications where clusters generate large volumes of data, but only a subset of this data or information extracted from it is needed to update their DTs. Cluster data is therefore transmitted to the associated MEC server for preprocessing, and the results are forwarded to the DT server.

Tasks are grouped into classes according to their preprocessing operation, input-data type, and freshness requirement. Operations may include compression, feature extraction, or semantic extraction, among others. Their parameters depend on the application and are not optimized in this work. Different classes may have different input sizes and computational demands, and each MEC server may receive a random number of tasks from these classes in each time slot. Freshness is measured using the AoI, which is tracked separately for each cluster and associated task class, as detailed in Section~\ref{sec:DT_fresh}.

The MEC servers allocate CPU resources to preprocessing tasks using a feasible set of strategies determined by the DT server. This set accounts for class-specific AoI requirements and the prescribed workload, delay, and update-generation bounds, as detailed in the next section.

A \emph{configuration}, indexed by $n$, is defined by its task classes, AoI requirements, and the statistical characteristics of task generation and workloads. Let $J_n$ and $K_n$ denote the numbers of task classes and feasible CPU-allocation strategies, respectively, and let $\epsilon_j^{(n)}$ denote the AoI requirement of class $j$. These quantities and the cluster associations remain fixed within a configuration. Task generation and workload statistics also remain stationary, although realized workloads vary and their distributions are unknown. A change in the defining characteristics initiates a new CPU-allocation problem and learning process. Here, we describe the system under a fixed configuration and omit its index when clear from the context, using $j\in[J]$ and $k\in[K]$ to index task classes and feasible strategies, respectively.

Within each configuration, we assume that all MEC servers have the same computing capacity $F>0$, measured in CPU cycles/s, dedicated to task preprocessing. The average total workload of each task class is also assumed to be the same across MEC servers, although realized workloads may differ across servers and time slots. These homogeneity assumptions allow the servers to collaboratively identify a common optimal CPU-allocation strategy, as described in Section~\ref{sec:problem_formulation}. With heterogeneous computing capacities or mean workloads, each MEC server could instead apply the framework independently, using its own feasible allocation set and local observations to identify its optimal CPU allocation.

\subsection{Task Generation and Computational Workload}
We consider time slots of duration $T>0$ to coordinate task generation and CPU allocation. Slots are indexed by $r=0,1,\ldots$. At the beginning of slot $r$, at time $t_r^{\mathrm G}=rT$, each cluster may generate tasks from the predefined task classes and offloads them to its associated MEC server.
Each task $q$ generated in slot $r$ contains data collected at its generation time, $t_q^{\mathrm G}=t_r^{\mathrm G}$. This timestamp is used to evaluate the freshness of the information incorporated into the DT.

Let $\mathcal Q_{m,j,r}$ be the set of tasks of class $j$ generated in slot $r$ and offloaded to MEC server $m$. Each task $q\in\mathcal Q_{m,j,r}$ has an input size of $\gamma_q$ bits and requires $W_q$ CPU cycles for preprocessing, where $W_q$ is a random variable with an unknown distribution. Its output size is $\nu_j\gamma_q$ bits, where $\nu_j\in(0,1]$ is an application-defined reduction factor that remains fixed for each class within a configuration.

The total input size and computational load of class $j$ at server $m$ in slot $r$ are
\begin{equation}
 \Gamma_{m,j,r}=\sum_{q\in\mathcal Q_{m,j,r}}\gamma_q,
 \qquad
 W_{m,j,r}=\sum_{q\in\mathcal Q_{m,j,r}}W_q.
 \label{eq:aggregate_workload}
\end{equation}
Both quantities are zero when $\mathcal Q_{m,j,r}=\phi$. 
The generation rate and size of the computational workloads are random, and their underlying distributions are unknown. However, we assume known bounds on the total workload and input size per class, server, and time slot:
\begin{equation}
 0\le W_{m,j,r}\le\overline W_j,  \qquad
 0\le \Gamma_{m,j,r}\le\overline \Gamma_j.
 \label{eq:workload_bounds}
\end{equation}
These bounds may follow, for example, from limits on the number of generated tasks, their input sizes, and the computational requirements of the tasks.

The CPU-allocation strategy determines the processing rate assigned to each task class and thereby the  tasks' execution times.  
\subsection{CPU-Allocation Strategies}

The MEC servers process the tasks using one of a finite set of $K$ feasible CPU-allocation strategies given by
   $ \{\boldsymbol f^{(1)},\ldots,\boldsymbol f^{(K)}\}$.
Strategy $k \in [K]$ specifies
\begin{equation}
 \boldsymbol f^{(k)}=[f_1^{(k)},\ldots,f_J^{(k)}]^\top,
 \quad f_j^{(k)}>0,
 \quad \sum_{j=1}^{J}f_j^{(k)}\le F.
 \label{eq:cpu_allocation}
\end{equation}
Here, $f_j^{(k)}$ denotes the share of the server's computing capacity allocated to class $j$, measured in Hz.

We assume that tasks within a class are processed sequentially, while different classes are processed concurrently. 
The allocated CPU rates remain unchanged regardless of which classes are present. Unused reservations are not redistributed to other classes during the slot.  Each server selects its strategy at the start of a slot, before observing the new task collection or workloads. The allocation remains fixed throughout the slot and may change only at the next slot boundary.

Motivated by resource-management architectures that separate performance monitoring from allocation decisions \cite{bhardwaj2023cilantro,wang2024autothrottle}, we consider a monitoring component at each MEC server that reports the total processing delay resulting from the CPU-strategy selection, without access to each class's execution time or computing capacity.
\subsection{Communication and Processing Delays}

For a task $q\in\mathcal Q_{m,j,r}$ generated by cluster $c$, the uplink transmission rate to MEC server $m$ is given by
\begin{equation}
 R_{q}^{\mathrm{UL}}(t_q^{\mathrm G})
 =  \mathscr{B}_{q,m}^{\mathrm{UL}}
 \log_2\left(
 1+\frac{P_{q}H_{c,m}(t_q^{\mathrm G})}
 {I_{c}(t_q^{\mathrm G})+I_{\mathrm N}^{2}}
 \right),
 \label{eq:cluster_mec_rate}
\end{equation}
where $\mathscr{B}_{q,m}^{\mathrm{UL}}$ is the allocated bandwidth, $P_q$ is the transmission power, $H_{c,m}$ is the channel power gain, and $I_c$ is the interference power affecting the cluster $c$'s transmission to its associated MEC receiver. The quantity $I_{\mathrm N}^{2}$ is the noise power over the allocated bandwidth. 

The uplink delay is $d_q^{\mathrm{UL}}=\mathrm{w}_q^{\mathrm{UL}}+\gamma_q/R_q^{\mathrm{UL}}$, where $\mathrm{w}_q^{\mathrm{UL}}\ge0$ is the waiting time for channel access and is zero when transmission can begin immediately. We assume that the uplink delay is bounded $d_q^{\mathrm{UL}}\le b_j^{\mathrm{UL}}$ for every task $q$ of class $j$. We assume a fixed upper bound \(u<T\) on the class uplink delay bounds:
\begin{equation}
    \max_{j\in[J]} b_j^{\mathrm{UL}}\le u<T.
\end{equation}
Tasks of class $j$ at MEC server $m$ are processed in their order of arrival. Let $q_{\text{old}}$ denote the task immediately preceding task $q$.
A task starts processing once it has arrived and the preceding task of the same class has completed processing. Assuming no unfinished workload carries over from the previous slot, the first task of each class starts processing upon arrival.
The processing start and completion times of task $q$ are given by
\begin{equation}
\begin{aligned}
 t_q^{\mathrm{Start},(k)}
 &= \max\left\{t_q^{\mathrm{Arr}},
              t_{q_{\text{old}}}^{\mathrm{Out},(k)}\right\},\\
 t_q^{\mathrm{Out},(k)}
 &= t_q^{\mathrm{Start},(k)}+\frac{W_q}{f_j^{(k)}}.
\end{aligned}
\label{eq:class_execution_time}
\end{equation}
Here, $t_q^{\mathrm{Arr}}$ is the arrival time at the MEC server, $W_q$ is the required computational workload in CPU cycles, and $f_j^{(k)}$ is the processing rate allocated to class $j$ under strategy $k$.

Under strategy $k$, the execution time of task $q$ is
\begin{equation}
    d_q^{\mathrm{proc},(k)}  =\frac{W_q}{f_j^{(k)}}.
 \label{eq:processing_delay}
\end{equation}
The total execution time of all class-$j$ tasks at server $m$  in slot $r$ is
\begin{equation}
    d_{m,j,r}^{\mathrm{exec},(k)}  =\sum_{q\in\mathcal Q_{m,j,r}}d_q^{\mathrm{proc},(k)}  =\frac{W_{m,j,r}}{f_j^{(k)}}.
    \label{eq:total_execution_time}
\end{equation}
This execution time is zero when no class-$j$ tasks arrive at server $m$ in slot $r$. Individual processing start and completion times follow
\eqref{eq:class_execution_time}.

Let $d_q^{\mathrm{DT},(k)}$ be the delay from completion of preprocessing until the result updates the DT. We assume that
\begin{equation}
    d_q^{\mathrm{DT},(k)} \leq b_j^{\mathrm{DT}},
    \label{eq:dt_delay_bound}
\end{equation}
for all class-$j$ tasks and candidate allocation strategies $k$. This assumes sufficient forwarding and DT-server capacity to ensure the bound under the modeled traffic conditions. The bound includes the transmission delay $\nu_j \gamma_q/R_q^{\mathrm{DT}}$, where $R_q^{\mathrm{DT}}$ is the forwarding rate, and any additional scheduling delays.

The task's DT update time and end-to-end delay are
\begin{equation}
 \begin{aligned}
    t_q^{\mathrm{DT},(k)}   &=t_q^{\mathrm{Out},(k)}+d_q^{\mathrm{DT},(k)},\\
    D_q^{(k)}  &=t_q^{\mathrm{DT},(k)}-t_r^{\mathrm G}\\
 &=d_q^{\mathrm{UL}} +\left(t_q^{\mathrm{Start},(k)}-t_q^{\mathrm {Arr}}\right)
   +\frac{W_q}{f_j^{(k)}} +d_q^{\mathrm{DT},(k)}.
 \end{aligned}
 \label{eq:end_to_end_delay}
\end{equation}
Here, $t_q^{\mathrm{Arr}}=t_r^{\mathrm G}+d_q^{\mathrm{UL}}$, and $t_q^{\mathrm{Start},(k)}-t_q^{\mathrm{Arr}}$ is the waiting time at the MEC server before execution begins.

All tasks generated in slot $r$ arrive by $t_r^{\mathrm G}+u$. The quantity $u$ is used only for delay analysis and does not impose a common processing start time. With no unfinished workload from the preceding slot, tasks of each class are processed sequentially at the reserved CPU rate as they become available. Since all tasks have arrived by $t_r^{\mathrm G}+u$, every class-$j$ task satisfies
\begin{equation}
    t_q^{\mathrm{Out},(k)}  \le t_r^{\mathrm G}+u+\frac{W_{m,j,r}}{f_j^{(k)}}.
\end{equation}
Consequently,
\begin{equation}
    D_q^{(k)}\le\overline D_j(k) :=u+\frac{\overline W_j}{f_j^{(k)}}+b_j^{\mathrm{DT}}.
    \label{eq:delay_bound}
\end{equation}
The feasible strategy set defined in the next section ensures that all tasks are processed and their results update the DT by the end of their generation slot. Hence, no unfinished workload is carried between slots.

\section{DT Information Freshness and Strategy Update}
\label{sec:DT_fresh}
A DT state is updated whenever a task is processed and its result is incorporated into the DT. We assume that each configuration change is accompanied by a fresh initialization of all monitored DT states. Thus, each configuration starts a new optimization problem with zero initial age.
\subsection{Digital-Twin Information Freshness}

For configuration $n$, let $\mathcal J_c^{(n)}\subseteq[J^{(n)}]$ denote the task classes associated with cluster $c$. Each pair $(c,j)$, with $j\in\mathcal J_c^{(n)}$, generates a sequence of updates to the corresponding DT state, with freshness requirement $\epsilon_j^{(n)}>0$. Updates belonging to the same class share the same freshness requirement, while freshness is tracked separately for each cluster–class pair.

Let $t_n^{\mathrm{cfg}}$ denote the beginning of  configuration $n$. At this time, all monitored DT states are initialized with fresh measurements generated at $t_n^{\mathrm{cfg}}$, such that
\begin{equation}
    A_{c,j}(t_n^{\mathrm{cfg}})=0, \qquad     \forall c\in\mathcal C,\;
    j\in\mathcal J_c^{(n)}.
\end{equation}
Let $\mathcal Q_{c,j}^{(n)}$ denote the tasks of class $j$ generated by cluster $c$ during configuration $n$. For $t\in[t_n^{\mathrm{cfg}},t_{n+1}^{\mathrm{cfg}})$, the generation time of the freshest information incorporated into the DT is
\begin{equation}
    U_{c,j}(t) = \max\left( \{t_n^{\mathrm{cfg}}\}
        \cup \left\{
            t_q^{\mathrm G}:
            q\in\mathcal Q_{c,j}^{(n)}, \; t_q^{\mathrm{DT}}\leq t
        \right\}
    \right).
    \label{eq:latest_generation_time}
\end{equation}
where $t_n^{\mathrm{cfg}}$ represents the timestamp of the freshly initialized DT state. For a task generated in slot $r$, $t_q^{\mathrm G}=rT$. Here, $t_q^{\mathrm{DT}}$ denotes the time at which the processed result updates the DT.

The age of information is
\begin{equation}
    A_{c,j}(t)=t-U_{c,j}(t).
    \label{eq:aoi}
\end{equation}

The age increases linearly over time and decreases when a result with a newer generation timestamp updates the DT. Results with equal or older timestamps do not reduce the age.

The normalized age is
\begin{equation}
    \widetilde A_{c,j}(t) = \frac{A_{c,j}(t)}{\epsilon_j^{(n)}}.
    \label{eq:normalized_aoi}
\end{equation}
%

\subsection{Freshness-Feasible CPU Allocations}
A CPU-allocation strategy is freshness-feasible if it allows the DT to incorporate updates while keeping the AoI within the corresponding freshness thresholds. A simple approach would be to require the delay bound in \eqref{eq:delay_bound} to satisfy $\overline{D}_j(k)\leq\epsilon_j$. However, this condition only considers the delay of individual updates and does not account for the time between consecutive generations. In practice, an update may not be generated in every slot, and some updates may be lost. We therefore assume that each cluster generates at least one valid update for each associated class $j$ in every $h_j$ consecutive slots that is successfully incorporated into the DT within the delay bound $\overline{D}_j(k)$, where $h_j\geq1$ is a known integer. This assumption holds for every candidate allocation strategy. With updates generated at slot boundaries, the maximum generation time of successfully delivered updates is $h_jT$. This permits slots with no new updates and allows other updates to be lost, while bounding the time between generations of successfully delivered updates.

Consider consecutive distinct generation slots $r_{\mathrm{prev}}$ and $r$ whose updates for cluster–class pair $(c,j)$ are successfully incorporated into the DT. Let $q$ be the first valid result from slot $r$ incorporated into the DT. Provided that the freshest incorporated update before this event was generated in slot $r_{\mathrm{prev}}$, the age immediately before incorporation is
\begin{equation}
 A_{c,j}\bigl((t_q^{\mathrm{DT}})^-\bigr)  =(r-r_{\mathrm{prev}})T+D_q,  \qquad r-r_{\mathrm{prev}}\le h_j,
 \label{eq:peak_aoi}
\end{equation}
where $D_q=t_q^{\mathrm{DT}}-t_r^{\mathrm G}$ is the actual end-to-end delay. The generation-slot gap is bounded by the assumption that every $h_j$ consecutive slots contain at least one generated update that is successfully delivered.

Using the conservative delay bound in \eqref{eq:delay_bound}, we construct a common set of $K$ freshness-feasible CPU-allocation strategies, indexed by $k\in[K]$. Every strategy satisfies
\begin{equation}
    \overline D_j(k)\le T, \quad  h_jT+\overline D_j(k)\le\epsilon_j,     \quad \forall j\in[J],\ \forall k\in[K].
    \label{eq:feasible_allocations}
\end{equation}
The first condition ensures that preprocessing finishes within the generation slot, preventing unfinished preprocessing workload from carrying into the next slot, and that successfully delivered updates are incorporated by that slot's end. The second condition ensures that the peak AoI remains within the freshness threshold despite slots with no generated updates or update losses, provided the successful-update assumption holds. For $h_j=1$, each cluster–class pair generates at least one successfully delivered update in every slot, and the freshness condition reduces to $T+\overline D_j(k)\le\epsilon_j$.

The DT constructs the candidate strategies using workload upper bounds and freshness requirements, without knowledge of the workload means or distributions. Specifically, the feasibility conditions in \eqref{eq:feasible_allocations} imply a minimum CPU allocation for each class:
\begin{equation}
    f_j^{\min} = \frac{\overline W_j}
    {\min\{T,\epsilon_j-h_jT\}-u-b_j^{\mathrm{DT}}}.
    \label{eq:minimum_cpu}
\end{equation}
We require positive denominators and
$\sum_{j=1}^{J}f_j^{\min}\le F$.
The remaining CPU capacity is distributed across classes to construct feasible candidates, let $v_j^{(k)}\ge0$ and $\sum_{j=1}^{J}v_j^{(k)}=1$, then 
\begin{equation}
    f_j^{(k)} = f_j^{\min}+\left(F-\sum_{\ell=1}^{J}f_\ell^{\min}\right)v_j^{(k)}.
    \label{eq:candidate_cpu}
\end{equation}
The weights $\boldsymbol v^{(k)}$ are selected to obtain diverse allocations across the feasible region without using expected losses. We assume that at least one feasible strategy exists, i.e., $K\geq1$; if $\sum_j f_j^{\min}=F$, the construction yields a single allocation. The feasible set is constructed before each configuration becomes effective and remains fixed throughout identification and deployment. Every strategy is available in every learning round, including slots in which some task classes are absent. The set and its size may change between configurations.

Under the stated initialization, update-generation, and delay assumptions, any sequence of feasible CPU-allocation strategies maintains $A_{c,j}(t)\leq\epsilon_j$ for every monitored cluster--class pair throughout the configuration. This freshness guarantee is independent of the statistical guarantee of best-arm identification. Without a finite bound on $h_j$, random task generation alone does not ensure a deterministic upper bound on AoI.

\section{Problem Formulation}
\label{sec:problem_formulation}

Within configuration $n$, all $K=K_n$ candidate CPU allocations satisfy the freshness requirements. However, these allocations distribute the computing capacity differently and may therefore produce different execution times for the same workload. Our objective is to identify a feasible allocation whose expected freshness-normalized total execution time is within a fixed tolerance of the minimum, with a predefined confidence and using as few observation slots as possible.

Each observation round corresponds to one time slot. Under allocation $k$, the total execution time of class-$j$ tasks at MEC server $m$ in slot $r$ is given by \eqref{eq:class_execution_time}.

We define the freshness-normalized execution time loss as
\begin{equation}
    L_{m,r}(k)
    = \sum_{j=1}^{J} \frac{d_{m,j,r}^{\mathrm{exec},(k)}}{\epsilon_j}
    = \sum_{j=1}^{J} \frac{W_{m,j,r}}{\epsilon_j f_j^{(k)}}.
    \label{eq:observed_loss}
\end{equation}
Normalization by $\epsilon_j$ assigns a larger loss to a given execution time for classes with tighter freshness requirements.

Within configuration $n$, we assume that the total class workloads have fixed conditional means. For every slot $r$ in this configuration,
\begin{equation}
    \mathbb E\!\left[ W_{m,j,r} \mid \mathcal F_{r-1},k_{m,r} \right]
    = \overline w_{m,j}^{(n)}, 
    \qquad m\in[M],\; j\in[J],
    \label{eq:workload_mean}
\end{equation}
where $\mathcal F_{r-1}$ contains the information available before slot $r$, and $k_{m,r}\in[K]$ is the index of the CPU-allocation strategy selected by MEC server $m$ before observing its new workloads.

To apply the distributed learning approach introduced later and satisfy its theoretical guarantees  \cite{yahya2026service}, we assume that these conditional means are common across MEC servers:
\begin{equation}
    \overline w_{m,j}^{(n)}  = \overline w_j^{(n)}, \qquad m\in[M],\; j\in[J].
    \label{eq:shared_workload_mean}
\end{equation}
This homogeneity assumption concerns the total mean workload of each class. It does not require identical workload realizations or distributions across servers.

Under \eqref{eq:workload_mean} and \eqref{eq:shared_workload_mean}, the expected execution time loss is common to all MEC servers and is given by
\begin{equation}
    \ell_k^{(n)} = \sum_{j=1}^{J} \frac{\overline w_j^{(n)}}{\epsilon_j f_j^{(k)}}.
    \label{eq:expected_loss}
\end{equation}
The optimal feasible allocation is
\begin{equation}
    k_n^* \in  \underset{k\in[K]}{\arg \min}  \, \ell_k^{(n)}.
    \label{eq:best_strategy}
\end{equation}
Minimizing \eqref{eq:expected_loss} selects the allocation with the smallest expected normalized execution time among these feasible candidates. This objective is different from the constraints in \eqref{eq:feasible_allocations}, which ensure that the AoI of every monitored cluster–class pair remains within its freshness threshold under the stated bounds.

We aim to identify and recommend an allocation $\hat k_n$ whose expected normalized execution time loss is within $\varepsilon_{\mathrm{BAI}}$ of the minimum with confidence at least $1-\delta$, using as few observation slots as possible. Specifically, the recommendation must satisfy
\begin{equation}
    \mathbb P\!\left[
        \ell_{\hat k_n}^{(n)}-\ell_{k_n^*}^{(n)}>\varepsilon_{\mathrm{BAI}}
    \right]\le\delta,
    \label{eq:pac}
\end{equation}
where $\varepsilon_{\mathrm{BAI}}=0$ corresponds to exact best-arm identification. This guarantee applies separately to each configuration. The per-agent failure probabilities $\delta_m$ are chosen such that $\delta = M\delta_m$.

\section{Linear Model for CPU Allocation}
\label{sec:structural_model}

In this section, we derive an exact linear representation of the expected execution time loss using features determined by the CPU allocations and freshness requirements, and describe the information observed at each MEC server. We note that alternative feature representations may also be considered, with linearity adopted as a modeling approximation rather than an exact relation, but these are not considered in this work.

\subsection{Linear Relation}

For a configuration $n$, we define the feature vector of allocation $k$ as
\begin{equation}
 \boldsymbol x_k  =  \left[
 \frac{1}{\epsilon_1 f_1^{(k)}},  \ldots,  \frac{1}{\epsilon_J f_J^{(k)}}  \right]^\top,
 \label{eq:structural_features}
\end{equation}
where the configuration index is omitted from the  features and allocations for notational simplicity. The unknown configuration-dependent parameter is
\begin{equation}
    \boldsymbol\theta_n^*  =  -\left[  \overline w_1^{(n)},\ldots,\overline w_J^{(n)}  \right]^\top.
 \label{eq:structural_parameter}
\end{equation}
Defining the expected reward as the negative expected execution time loss, we obtain
\begin{equation}
 \mu_k^{(n)}  := -\ell_k^{(n)} = \boldsymbol x_k^\top\boldsymbol\theta_n^*.
 \label{eq:structural_linearity}
\end{equation}
This relation is exact under the stated execution and workload assumptions. The features depend only on the predefined CPU allocations and freshness thresholds and remain fixed throughout the configuration. 

\subsection{Scalar Feedback}

Following resource-management architectures that separate performance monitoring from allocation decisions \cite{bhardwaj2023cilantro,wang2024autothrottle}, we consider a monitoring component at each MEC server that normalizes task execution times by the class-specific freshness thresholds and sums them over each slot. At the end of the slot, this component reports the resulting scalar execution time loss to the allocation learner. We assume that the learner receives only this aggregate feedback, without access to execution times or realized workloads.

For configuration $n$, let $\tau_n=r_nT$ denote the start of its identification phase, where $T$ is the slot duration and $r_n$ is the first observation slot. We index consecutive observation rounds by $s=1,2,\ldots$, with round $s$ corresponding to slot $r=r_n+s-1$ and beginning at time $\tau_n+(s-1)T$. Each round therefore lasts one slot. Server $m$ selects $k_{m,n,s}=k_{m,r}\in[K]$ before observing the slot's workload and receives the scalar reward
\begin{equation}
    \begin{aligned}
        Y_{m,n,s}
        &=-L_{m,r}(k_{m,n,s})\\
        &=\boldsymbol{x}_{k_{m,n,s}}^\top
          \boldsymbol{\theta}_n^*+\eta_{m,n,s},
    \end{aligned}
    \label{eq:linear_observation}
\end{equation}
where
\begin{equation}
    \eta_{m,n,s}
    =-\sum_{j=1}^{J}
    \frac{W_{m,j,r}-\overline w_j^{(n)}}
    {\epsilon_j f_j^{(k_{m,n,s})}}
    \label{eq:observation_noise}
\end{equation}
is the deviation of the realized workloads from their conditional means.

Let $\boldsymbol{k}_r=(k_{1,r},\ldots,k_{M,r})$ be the allocations selected by all MEC servers in slot $r$. We assume that the shared conditional workload means remain $\overline w_j^{(n)}$ when conditioned on $\mathcal F_{r-1}$ and $\boldsymbol{k}_r$. Consequently,
\begin{equation}
   \mathbb E\!\left[
        \eta_{m,n,s}\mid\mathcal F_{r-1},\boldsymbol{k}_r
    \right]=0.
    \label{eq:zero_mean_noise}
\end{equation}
%
\subsection{Observation and Parameter Bounds}

Since $0\le W_{m,j,r}\le\overline W_j$, define
\begin{equation}
    L_{\max}
    =\max_{k\in[K]}
    \sum_{j=1}^{J}
    \frac{\overline W_j}{\epsilon_j f_j^{(k)}}.
    \label{eq:loss_bound}
\end{equation}
Then $Y_{m,n,s}\in[-L_{\max},0]$. Together with the conditional zero-mean property in
\eqref{eq:zero_mean_noise}, Hoeffding's lemma gives
\cite{lattimore2020bandit}
\begin{equation}
    \mathbb E\!\left[
        e^{\alpha\eta_{m,n,s}}
        \mid\mathcal F_{r-1},\boldsymbol{k}_r
    \right]
    \le
    \exp\left(\frac{\alpha^2\sigma^2}{2}\right),
    \sigma=\frac{L_{\max}}{2}, \alpha\in\mathbb R.
    \label{eq:subgaussian_noise}
\end{equation}
Thus, the observation noise is conditionally $\sigma$-sub-Gaussian.
The workload bounds also imply the parameter-norm bound
\begin{equation}
    \|\boldsymbol{\theta}_n^*\|_2
    \le
    S:=
    \left(\sum_{j=1}^{J}\overline W_j^2\right)^{1/2}.
    \label{eq:parameter_norm_bound}
\end{equation}
For the finite candidate set, a feature-norm bound is
\begin{equation}
    \|\boldsymbol{x}_k\|_2
    \le
    \max_{k'\in[K]}
    \left(
        \sum_{j=1}^{J}
        \frac{1}{\epsilon_j^2(f_j^{(k')})^2}
    \right)^{1/2}.
    \label{eq:feature_norm_bound}
\end{equation}

These bounds are determined for each configuration and remain fixed throughout its identification phase. They are used to construct the confidence bounds in the distributed identification method introduced next.

\section{Distributed Bandit Learning for CPU Allocation}
\label{sec:collaborative_learning}

In this section, we formulate MEC CPU-allocation selection as a fixed-confidence best-arm identification problem in linear bandits. The objective is to identify an allocation with the predefined accuracy and confidence using as few learning rounds as possible. Each MEC server can solve this problem independently using LinGapE \cite{xu2018fully}, without requiring workload homogeneity across servers. Under the homogeneity assumptions introduced above, the servers can instead share observations to identify a common allocation. We adopt DistLinGapE \cite{yahya2026service} to exploit this collaboration and accelerate identification.

Each MEC server acts as an agent, and each feasible CPU-allocation strategy $k\in[K]$ corresponds to an arm.
Each configuration defines a separate identification problem, with observation rounds corresponding to the learning rounds of that problem. If identification uses $N_n$ consecutive observation slots, it ends at time
$\hat\tau_n=\tau_n+N_nT$.

Before identification begins, the DT determines the feasible candidate allocations using the prescribed freshness requirements and workload, delay, and update-generation bounds. During each learning round, every server selects an allocation, applies it to the arriving tasks, and observes the scalar reward defined in \eqref{eq:linear_observation}. Meanwhile, the processed results are forwarded to the DT. The reward from each round is available before the allocation decision for the following round.
We describe the procedure within a fixed configuration and omit the configuration index $n$ for readability. 

If server $m$ were to learn independently, its local observations would give the following quantities at the beginning of learning round $s$, with the server index omitted for simplicity:
\begin{equation}
    \boldsymbol{A}_s =\lambda\boldsymbol{I}  +\sum_{z=1}^{s-1} \boldsymbol{x}_{k_z}\boldsymbol{x}_{k_z}^{\top},
    \quad
    \boldsymbol{b}_s =\sum_{z=1}^{s-1}\boldsymbol{x}_{k_z}Y_z,
\end{equation}
where $\lambda>0$ is the regularization parameter,
$\boldsymbol{I}$ is the $J\times J$ identity matrix,
$k_z$ is the arm selected in round $z$, and $Y_z$ is the corresponding observed reward. The regularized least-squares estimate of the parameter vector would be $\hat{\boldsymbol{\theta}}_s=\boldsymbol{A}_s^{-1}\boldsymbol{b}_s$ \cite{xu2018fully}.

However, When the servers collaborate, each server has access to its local observations in addition to information collected by the other servers. A coordinator, located at either an MEC server or the DT server, aggregates and distributes this information.

Algorithm~\ref{alg:DistLinGapE} on the next page and the following description summarize the application of DistLinGapE to our CPU-allocation problem. The detailed algorithm and theoretical analysis are provided in \cite{yahya2026service}.
\subsection{Local Estimation and Information Sharing}

Let $s_\ell$ denote the first learning round of communication round $\ell$, with $s_1=1$. A communication round comprises the learning rounds between two successive synchronizations. At the beginning of round $s$, server $m$ stores the statistics of its observations collected since round $s_\ell$:
\begin{equation}
\begin{aligned}
    \Delta\boldsymbol{A}_{m,s}
    &= \sum_{z=s_\ell}^{s-1}
    \boldsymbol{x}_{k_{m,z}}\boldsymbol{x}_{k_{m,z}}^\top,\\
    \Delta\boldsymbol{b}_{m,s}
    &= \sum_{z=s_\ell}^{s-1}
    \boldsymbol{x}_{k_{m,z}}Y_{m,z}.
\end{aligned}
\label{eq:local_statistics}
\end{equation}
Here, $k_{m,z}$ and $Y_{m,z}$ are the allocation selected by server $m$ and its reward in round $i$, respectively.

Let $\boldsymbol{A}_{\textrm{coor},\ell}$ and $\boldsymbol{b}_{\textrm{coor},\ell}$ denote the cumulative statistics shared at the beginning of communication round $\ell$, initially both zero. Each server combines these statistics with its unshared local observations:
\begin{equation}
\begin{aligned}
    \boldsymbol{A}_{m,s}
    &= \lambda\boldsymbol{I}
    +\boldsymbol{A}_{\textrm{coor},\ell}
    +\Delta\boldsymbol{A}_{m,s},\\
    \boldsymbol{b}_{m,s}
    &= \boldsymbol{b}_{\textrm{coor},\ell}
    +\Delta\boldsymbol{b}_{m,s},\\
    \hat{\boldsymbol{\theta}}_{m,s}
    &= \boldsymbol{A}_{m,s}^{-1}\boldsymbol{b}_{m,s}.
\end{aligned}
\label{eq:theta_est}
\end{equation}
Thus, each estimate is based on the total observations from all servers up to the latest synchronization, together with the server's subsequent local observations.

Communication and information sharing follows the event-triggered rule of \cite{yahya2026service}, which measures the reduction in the volume of the confidence ellipsoid for the estimate of $\boldsymbol{\theta}^{*}$ since the beginning of the current communication round. When any server meets this condition, all servers transmit their unshared statistics to the coordinator. If communication is triggered after learning round $s$, the coordinator updates
\begin{equation}
    \begin{aligned}
    \boldsymbol{A}_{\textrm{coor},\ell+1}
    &= \boldsymbol{A}_{\textrm{coor},\ell}
    +\sum_{m=1}^{M}\Delta\boldsymbol{A}_{m,s+1},\\
    \boldsymbol{b}_{\textrm{coor},\ell+1}
    &= \boldsymbol{b}_{\textrm{coor},\ell}
    +\sum_{m=1}^{M}\Delta\boldsymbol{b}_{m,s+1},
    \end{aligned}
\label{eq:coor_stat}
\end{equation}
where the increments include all observations collected since the previous synchronization until round $s$. The coordinator broadcasts its updates, and the servers clear their local increments. The next communication round starts at $s_{\ell+1}=s+1$.

\subsection{Allocation Selection and Stopping Rule}

Each server selects CPU allocations sequentially to identify the best feasible allocation. In each learning round, it selects an allocation that helps distinguish its currently preferred allocation from a competing allocation whose performance remains uncertain. Under the noise and parameter assumptions of \cite{yahya2026service}, define the confidence radius for the parameter estimate \cite{abbasi2011improved,yahya2026service}
\begin{equation}
    C_{m,s}
    =\sigma\sqrt{2\log\left(
    \frac{\det(\boldsymbol{A}_{m,s})^{1/2}}
    {\lambda^{J/2}(\delta/M)}
    \right)}
    +\sqrt{\lambda}S,
\label{eq:conf_reward}
\end{equation}
where $\sigma$ is the conditional sub-Gaussian noise parameter, $S$ bounds $\|\boldsymbol{\theta}^{*}\|_2$, and $\delta$ is the fixed error probability.

For allocations $i$ and $j$, the estimated reward difference and its confidence radius are \cite{yahya2026service}
\begin{equation}
\begin{aligned}
    \hat{\Delta}_{m,s}(i,j)
    &=(\boldsymbol{x}_i-\boldsymbol{x}_j)^\top
    \hat{\boldsymbol{\theta}}_{m,s},\\
    \beta_{m,s}(i,j)
    &=C_{m,s}
    \|\boldsymbol{x}_i-\boldsymbol{x}_j\|_{\boldsymbol{A}_{m,s}^{-1}}.
\end{aligned}
\label{eq:gap_est_bound}
\end{equation}
Server $m$ first identifies the allocation with the largest estimated reward:
\begin{equation}
    i_{m,s}\in  \underset{i\in[K]}{\arg\max} \,\,
    \boldsymbol{x}_i^\top\hat{\boldsymbol{\theta}}_{m,s}.
\label{eq:local_best}
\end{equation}
For $K\geq2$, it then selects the most ambiguous allocation, i.e., the one with the largest upper confidence bound on its improvement over $i_{m,s}$:
\begin{equation}
    j_{m,s}\in
    \underset{j\in[K]\setminus\{i_{m,s}\}}{\arg\max} 
    \left\{
    \hat{\Delta}_{m,s}(j,i_{m,s})
    +\beta_{m,s}(j,i_{m,s})
    \right\}.
\label{eq:local_challenger}
\end{equation}
The corresponding bound is
\begin{equation}
    B_{m,s}
    =\hat{\Delta}_{m,s}(j_{m,s},i_{m,s})
    +\beta_{m,s}(j_{m,s},i_{m,s}).
\label{eq:stopping_bound}
\end{equation}
If $B_{m,s}\leq\varepsilon_{\mathrm{BAI}}$, server $m$ reports $i_{m,s}$ to the coordinator and identification terminates. Here, $\varepsilon_{\mathrm{BAI}}\geq0$ is the permitted reward suboptimality. Under the assumptions of \cite{yahya2026service}, the returned allocation $\hat k$ satisfies
\begin{equation}
    \mathbb{P} \left(
    \max_{k\in[K]}\boldsymbol{x}_k^\top\boldsymbol{\theta}^{*}
    -\boldsymbol{x}_{\hat k}^\top\boldsymbol{\theta}^{*}
    \leq\varepsilon_{\mathrm{BAI}}
    \right)
    \geq1-\delta.
\label{eq:distlingape_guarantee}
\end{equation}

If the stopping condition is not met, the server samples an allocation that helps distinguish $i_{m,s}$ from $j_{m,s}$. DistLinGapE supports greedy sampling and allocation-ratio tracking
\cite{yahya2026service,xu2018fully}. The greedy rule selects
\begin{equation}
    k_{m,s}\in  \underset{k\in[K]}{\arg\min} \,
    \|\boldsymbol{x}_{i_{m,s}}-\boldsymbol{x}_{j_{m,s}}\|_
    {(\boldsymbol{A}_{m,s}+\boldsymbol{x}_k\boldsymbol{x}_k^\top)^{-1}}.
\label{eq:arm_greedy}
\end{equation}
This rule selects the allocation that minimizes the uncertainty for the estimated reward difference between $i_{m,s}$ and $j_{m,s}$ after observing its reward. Any candidate allocation may be selected, not only $i_{m,s}$ and $j_{m,s}$, since the linear model allows observations from other allocations to inform this reward difference.

Alternatively, allocation-ratio tracking selects allocations according to target sampling proportions to distinguish between $i_{m,s}$ and $j_{m,s}$. The target proportions and tracking rule are given in \cite{yahya2026service,xu2018fully}.

After identification, the coordinator distributes the recommended CPU allocation to the MEC servers for deployment.

\begin{algorithm}[t] 
\caption{DistLinGapE for CPU-Allocation Identification:
    Summary of \cite{yahya2026service}}
\label{alg:DistLinGapE}
\begin{algorithmic}[1]
\STATE \textbf{Input:}
$\{\boldsymbol{x}_k\}_{k=1}^{K}$,
$\varepsilon_{\mathrm{BAI}}$, $\delta$, $\lambda$,
$\sigma$, $S$, and communication threshold $D_{\text{th}}$.
\STATE Initialize shared and local statistics to zero;
set $\ell=1$ and $s_1=1$.
\FOR{$s=1,2,\ldots$}
    \STATE Each server forms its parameter estimate
    using \eqref{eq:theta_est}.
    \STATE Each server computes $i_{m,s}$, $j_{m,s}$,
    and $B_{m,s}$ using
    \eqref{eq:local_best}--\eqref{eq:stopping_bound}.
    \IF{some server $m$ has
    $B_{m,s}\leq\varepsilon_{\mathrm{BAI}}$}
        \STATE Broadcast and \textbf{return}
        $\hat k=i_{m,s}$.
    \ENDIF
    \STATE Each server selects an allocation $k_{m,s}$ using
    \eqref{eq:arm_greedy} or allocation-ratio tracking
    \cite{yahya2026service}.
    \STATE Each server applies its allocation in
    the slot corresponding to round $s$, observes $Y_{m,s}$,
    and updates its local statistics.
    \IF{any server meets the communication condition
    with threshold $D_{\text{th}}$ \cite{yahya2026service}}
        \STATE All servers transmit their unshared updates to the coordinator, which aggregates them using \eqref{eq:coor_stat} and broadcasts the updated totals. 
        \STATE Servers clear local increments and reset the
        communication counters; set
        $s_{\ell+1}=s+1$ and $\ell\leftarrow\ell+1$.
    \ENDIF
\ENDFOR
\end{algorithmic}
\end{algorithm}

\section{Numerical Results}
\label{sec:numerical_results}

In this section, we show the performance of our proposed approach for identifying a CPU-allocation strategy with minimum expected execution delay normalized by freshness, while meeting the task classes' freshness requirements.
\subsection{Simulation Setup}

We consider a $600\,\mathrm{m}\times600\,\mathrm{m}$ region containing $M=4$ MEC servers and $C=16$ IoT clusters. The servers are located at $(150,150)$, $(150,450)$, $(450,150)$, and $(450,450)$ m. Four clusters are independently and uniformly placed within each server's quadrant and associated with their nearest MEC server. Each server has a processing capacity of $F=8$ GHz and processes tasks belonging to $J=8$ classes. The slot duration is $T=200$ ms.

At the beginning of slot $r$, cluster $c$ generates $N_{c,j,r}=1+\operatorname{Bernoulli}(p_j)$ tasks of class $j$. In the main experiment, we use
$\boldsymbol p=[0.95,0.90,0.60,0.50,0.20,0.20,0.10,0.10]^\top$.
Each task has an independently and uniformly distributed size between 40 and 60 kbytes. The class-specific computational requirements are
$\boldsymbol\kappa=[44,40,20,16,6,5,4,3]^\top$
cycles per input bit.

Each cluster has an uplink rate of 30 Mbit/s, with the four clusters associated with a server served simultaneously. Uploading the maximum total input generated by one cluster requires 17.68 ms. We therefore use a common uplink delay bound of $u=b_j^{\mathrm{UL}}=20$ ms. Each MEC server has a dedicated 100 Mbit/s connection to the DT server, and preprocessing reduces the forwarded data volume to $10\%$ of the input size. The combined forwarding and DT-incorporation delay is assumed to be bounded by $b_j^{\mathrm{DT}}=1$ ms.

The class-specific freshness requirements are
$\boldsymbol\epsilon=[400,400,400,400,375,375,350,350]^\top\ \mathrm{ms}$.
Since every cluster generates at least one task of each class in every slot, we use $h_j=1$.

\subsection{Candidate Strategies and Evaluation}

Each candidate set contains $K=30$ distinct feasible CPU-allocation strategies. We determine the minimum allocations from the workload upper bounds and freshness requirements using \eqref{eq:minimum_cpu}, and distribute the remaining capacity according to \eqref{eq:candidate_cpu}. Every set includes equal CPU allocation, $f_j^{\mathrm{eq}}=1$ GHz, and an allocation that distributes the residual capacity equally above the class-specific minima.

For the main experiment, the remaining candidates comprise 14 allocations generated using symmetric $\operatorname{Dirichlet}(1,\ldots,1)$ weights and 14 using $\operatorname{Dirichlet}(3,\ldots,3)$ weights.
The results are obtained over 100 algorithm runs.

The suboptimality gap of candidate $k$ is
$\Delta_k=\mu_{k^\star}-\mu_k$.
The smallest gap,
$\Delta_{\min}=\min_{k\ne k^\star}\Delta_k$,
is the expected reward difference between the best and second-best candidates. In the main example, $\Delta_{\min}\approx0.0652$. We also consider candidate sets with broader and smaller overall reward separation and minimum gaps in Figs~\ref{fig:delay_large_gaps} and~\ref{fig:delay_small_gaps}.

We compare DistLinGapE with equal CPU allocation, random allocation, independent LinGapE, shared uniform exploration, and an oracle. Equal CPU allocation always assigns the same capacity to every class, whereas random allocation selects uniformly from the candidate set in each round. Independent LinGapE uses only local observations, and each server stops and deploys its recommendation independently. Shared uniform exploration cycles through all candidates equally, while using the same linear estimator, communication rule, and stopping criterion as DistLinGapE. The oracle knows the mean workloads and selects the candidate minimizing the expected freshness-normalized execution time loss.

Each server receives one scalar observation per round, giving $M$ observations across the network. The learning methods begin with 30 initialization rounds in which each server samples every candidate once. We use an identification tolerance of $\varepsilon_{\mathrm{BAI}}=0.02$, an overall error probability of $\delta=0.2$, and ridge regularization $\lambda=0.1$. Learners observe only the scalar freshness-normalized execution time loss. After identification, each method deploys its recommended allocation for the remaining rounds.

\subsection{Freshness-Normalized Loss and CPU Execution Delay}

We first evaluate the freshness-normalized execution time loss averaged across MEC servers,
\begin{equation*}
    \frac{1}{M}\sum_{m=1}^{M}L_{m,r}(k_{m,r})
    =
    \frac{1}{M}\sum_{m=1}^{M}\sum_{j=1}^{J}
    \frac{W_{m,j,r}}{\epsilon_j f_j^{(k_{m,r})}}.
\end{equation*}
This dimensionless metric directly corresponds to the optimization objective.

\begin{figure}
    \centering
    \includegraphics[width=0.9\linewidth]
    {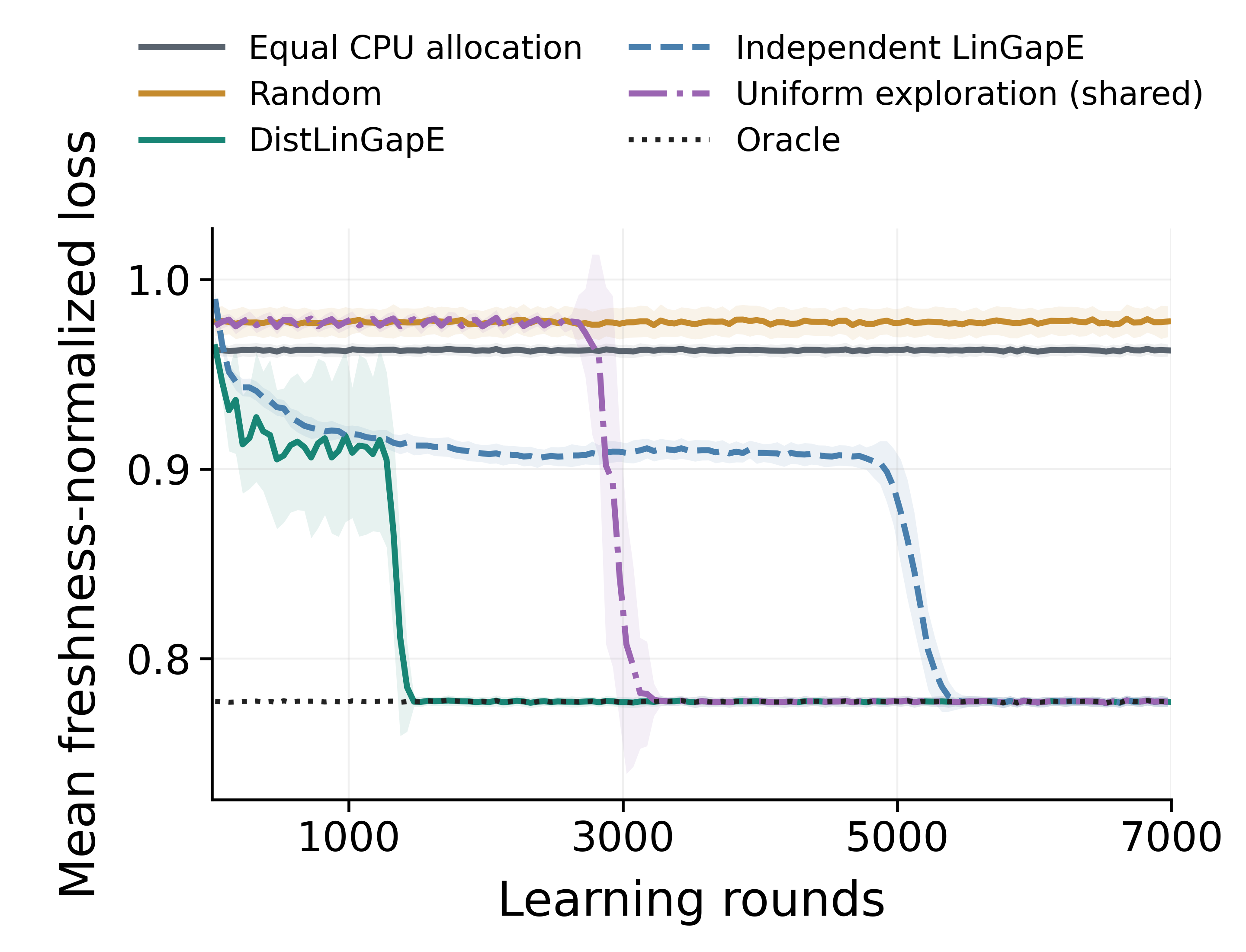}
    \caption{Mean freshness-normalized execution time loss for the main candidate set.}
    \label{fig:normalized_moderate_gaps}
\end{figure}

Fig.~\ref{fig:normalized_moderate_gaps} shows that DistLinGapE reaches the oracle's performance earlier than the other learning methods. It satisfies its stopping condition after an average of 1354 rounds, with stopping times ranging from 1260 to 1455 rounds across the 100 trials. Shared uniform exploration requires an average of 2932 rounds. Independent LinGapE requires 5127 rounds per server on average, while all four servers complete identification after an average of 5264 rounds.

All three learning methods identify the exact best candidate in this experiment. Over the final 500 rounds, their mean normalized losses coincide with the oracle at approximately 0.7774, compared with 0.9627 for equal CPU allocation and 0.9776 for random allocation. The benefit of DistLinGapE is therefore earlier identification and deployment, rather than a different final allocation.

To express the resulting performance in milliseconds, we also show the unnormalized mean class CPU execution time,
\begin{equation*}
    \frac{1}{MJ}
    \sum_{m=1}^{M}\sum_{j=1}^{J}
    \frac{W_{m,j,r}}{f_j^{(k_{m,r})}}.
\end{equation*}
This metric averages class execution durations across servers and is distinct from per-task delay and end-to-end DT update delay. Minimizing the freshness-normalized loss does not generally minimize this unnormalized metric, since classes receive different weights through their freshness requirements. Here, the thresholds are relatively close, ranging from 350 to 400 ms, so the weights differ only modestly. We therefore report CPU execution times as an additional practical performance measure, while assessing optimality using the normalized objective.

\begin{figure}
    \centering
    \includegraphics[width=0.9\linewidth]
    {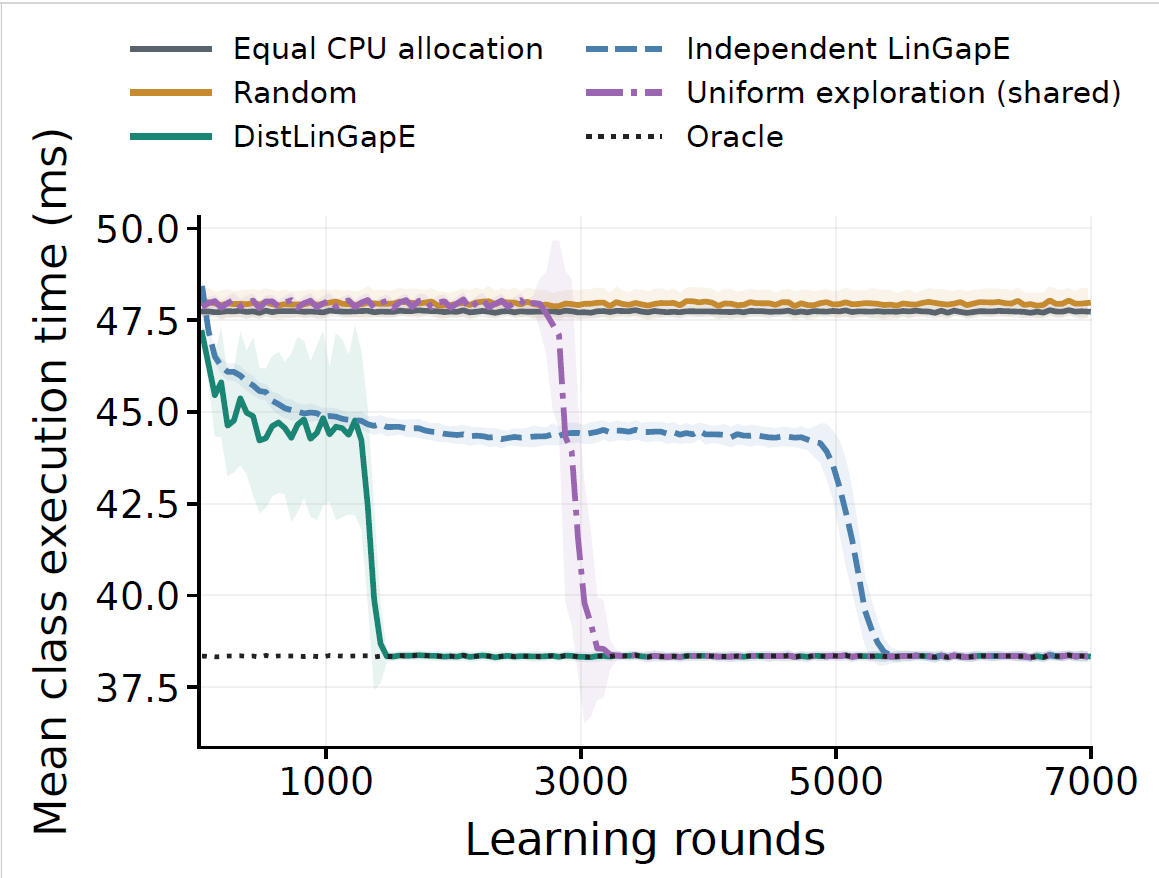}
    \caption{Mean class CPU execution time for the main candidate set.}
    \label{fig:delay_moderate_gaps}
\end{figure}

Fig.~\ref{fig:delay_moderate_gaps} shows that the allocation selected by all three learning methods and the oracle achieves a mean execution time of approximately 38.34 ms over the final 500 rounds, compared with 47.74 ms for equal CPU allocation and 47.96 ms for random allocation. Thus, optimizing the normalized objective also reduces the unnormalized execution time in this experiment.

\begin{figure}
    \centering
    \includegraphics[width=0.9\linewidth]
    {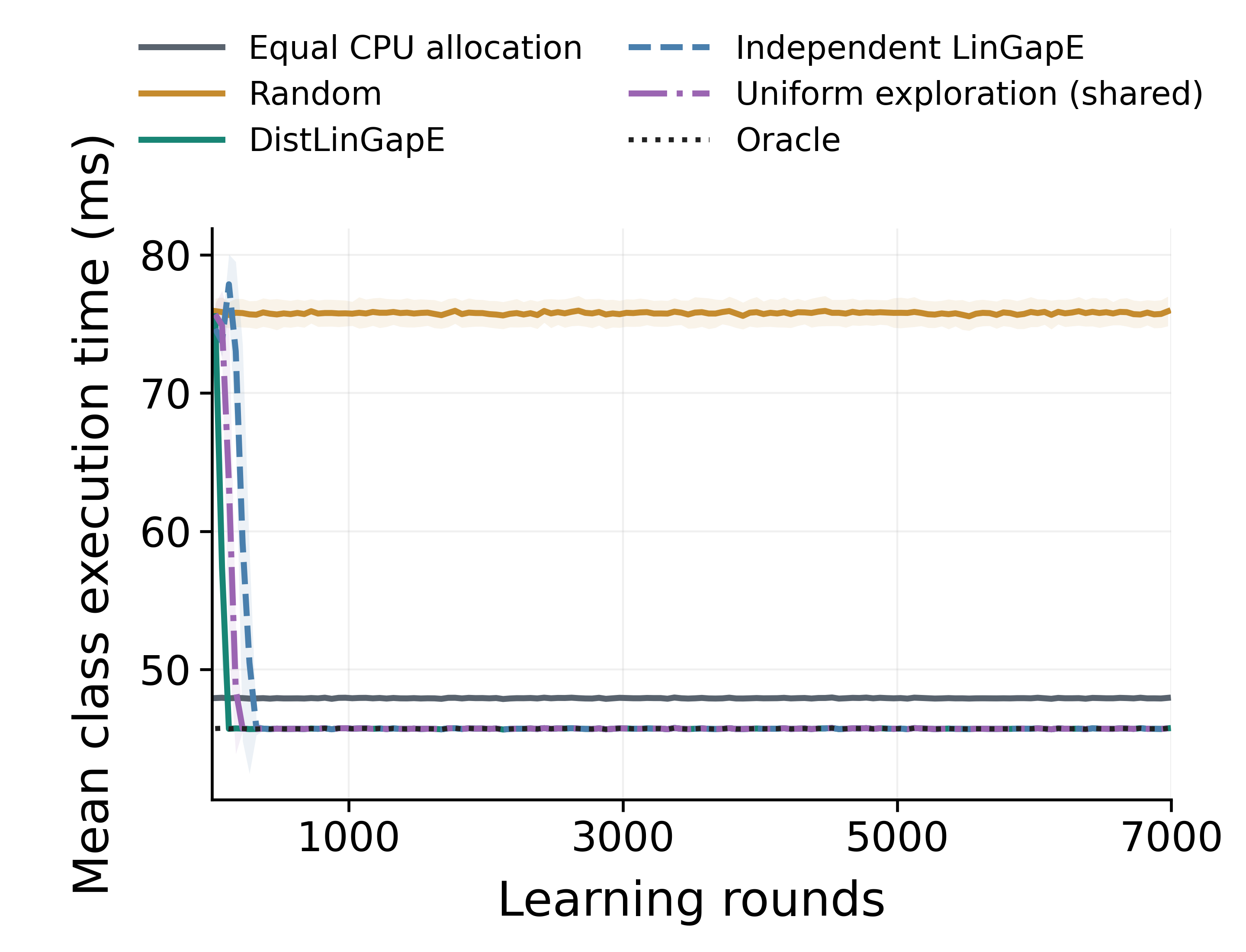}
    \caption{Mean class CPU execution time for geometrically spread feasible candidates.}
    \label{fig:delay_large_gaps}
\end{figure}

\begin{figure}
    \centering
    \includegraphics[width=0.9\linewidth]
    {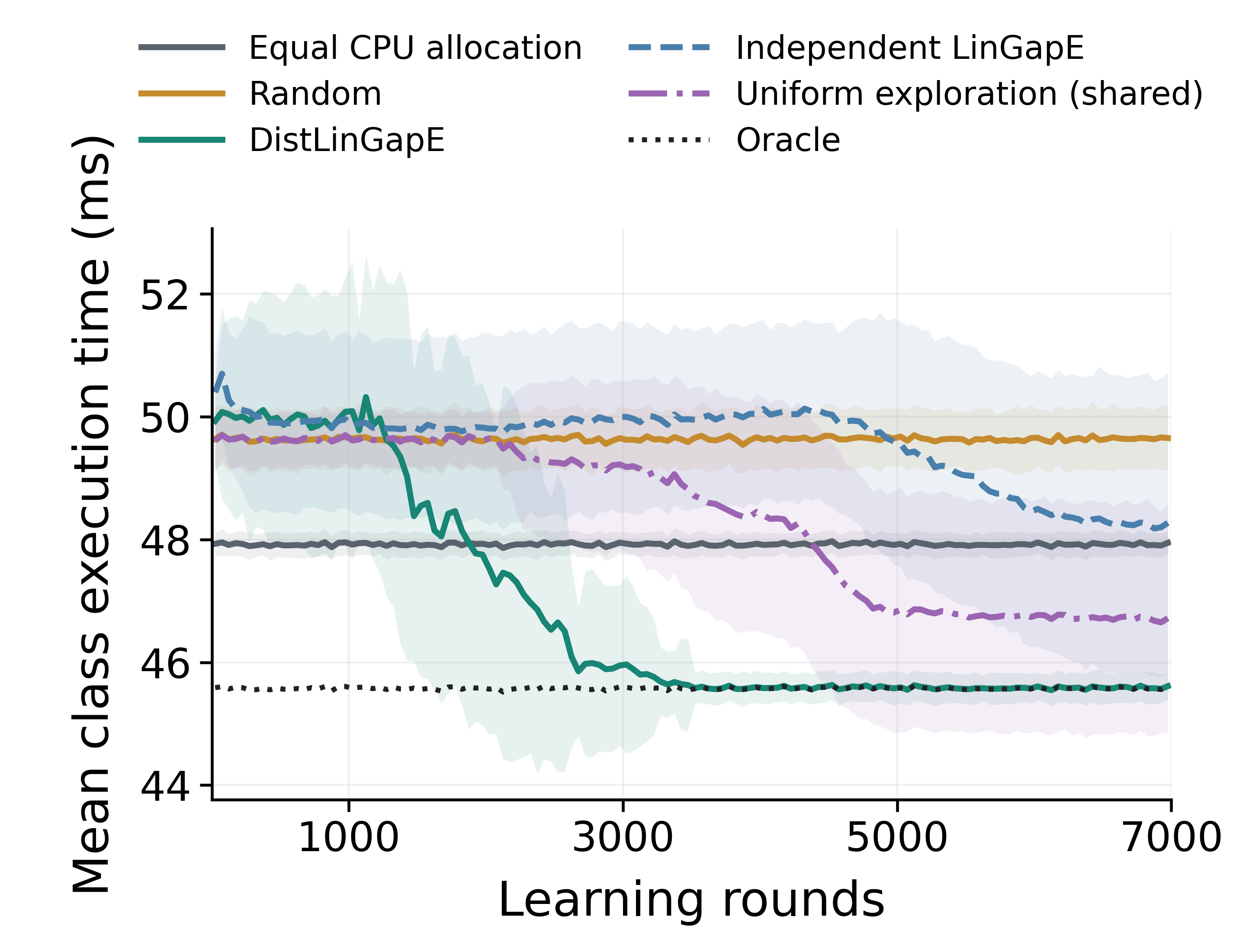}
    \caption{Mean class CPU execution time for candidate sets with small best--second-best reward gaps.}
    \label{fig:delay_small_gaps}
\end{figure}

We also evaluate allocation sets with broader or smaller overall reward separation and minimum gaps in Figs.~\ref{fig:delay_large_gaps} and~\ref{fig:delay_small_gaps}. For these experiments, we use
$\boldsymbol p=[0.8,0.7,0.6,0.5,0.4,0.3,0.2,0.1]^\top$
and computation requirements
$\boldsymbol\kappa=[32,28,24,20,16,14,12,10]^\top$
cycles per input bit. Each experiment includes ten candidate sets and ten independent workload realizations per set. The freshness thresholds remain unchanged.

Fig.~\ref{fig:delay_large_gaps} considers candidates selected by geometric coverage from options generated using symmetric Dirichlet distributions with concentrations 0.5, 1, and 5. This produces larger reward gaps to many suboptimal candidates. DistLinGapE stops after an average of 68 rounds, compared with 133 rounds for shared uniform exploration and 221 rounds per server for independent LinGapE. All three methods identify the exact best candidate under the normalized objective in all trials.

Fig.~\ref{fig:delay_small_gaps} considers candidates generated using symmetric Dirichlet distributions with concentrations 2 and 5, without separation filtering. Their best--second-best reward gaps range from approximately $6.43\times10^{-5}$ to $0.0124$. DistLinGapE requires an average of 1948 rounds. All 100 recommendations are $\varepsilon_{\mathrm{BAI}}$-optimal for $\varepsilon_{\mathrm{BAI}}=0.02$, while 89 identify the exact best candidate.

Within the 7000-round horizon, shared uniform exploration completes identification in 72 of the 100 trials. For independent LinGapE, all four servers finish in 40 of the 100 trials. Their late-stage execution times therefore include continued exploration in unfinished runs.

These results show that identification complexity depends on the observed rewards, reward gaps, and the stopping conditions. DistLinGapE uses adaptive sampling and information sharing to reduce the exploration needed before deployment. Since its objective is identification rather than immediate delay minimization, it may select suboptimal allocations during exploration.

\subsection{Collaboration Speedup}

We evaluate the benefit of collaboration using the small-gap candidate sets and $M\in\{1,2,4,8\}$ agents. Each server serves four clusters, giving $C=4M$. This preserves the per-agent workload distribution while increasing the number of observations collected concurrently. For $M=1$, the agent learns independently without communication. Each value of $M$ is evaluated over 100 trials using the same candidate sets and fresh workload realizations.

\begin{figure}
    \centering
    \includegraphics[width=0.7\linewidth]
    {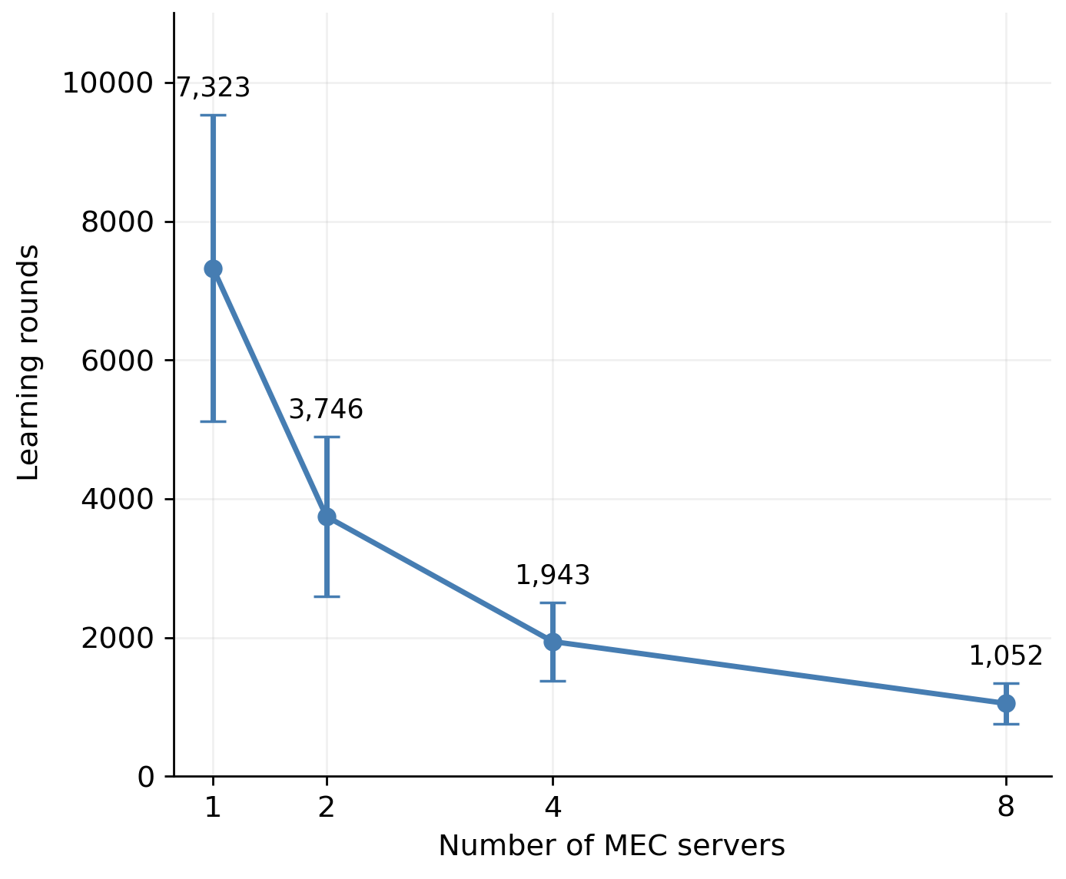}
    \caption{Mean identification rounds versus the number of collaborating MEC servers for the small-gap candidate sets.}
    \label{fig:small_gap_rounds_speedup}
\end{figure}

Fig.~\ref{fig:small_gap_rounds_speedup} shows that the mean number of identification rounds decreases from approximately 7323 for one agent to 3746, 1943, and 1052 for two, four, and eight agents, respectively. All runs satisfy the stopping condition and return an $\varepsilon_{\mathrm{BAI}}$-optimal candidate.

\begin{figure}
    \centering
    \includegraphics[width=0.7\linewidth]
    {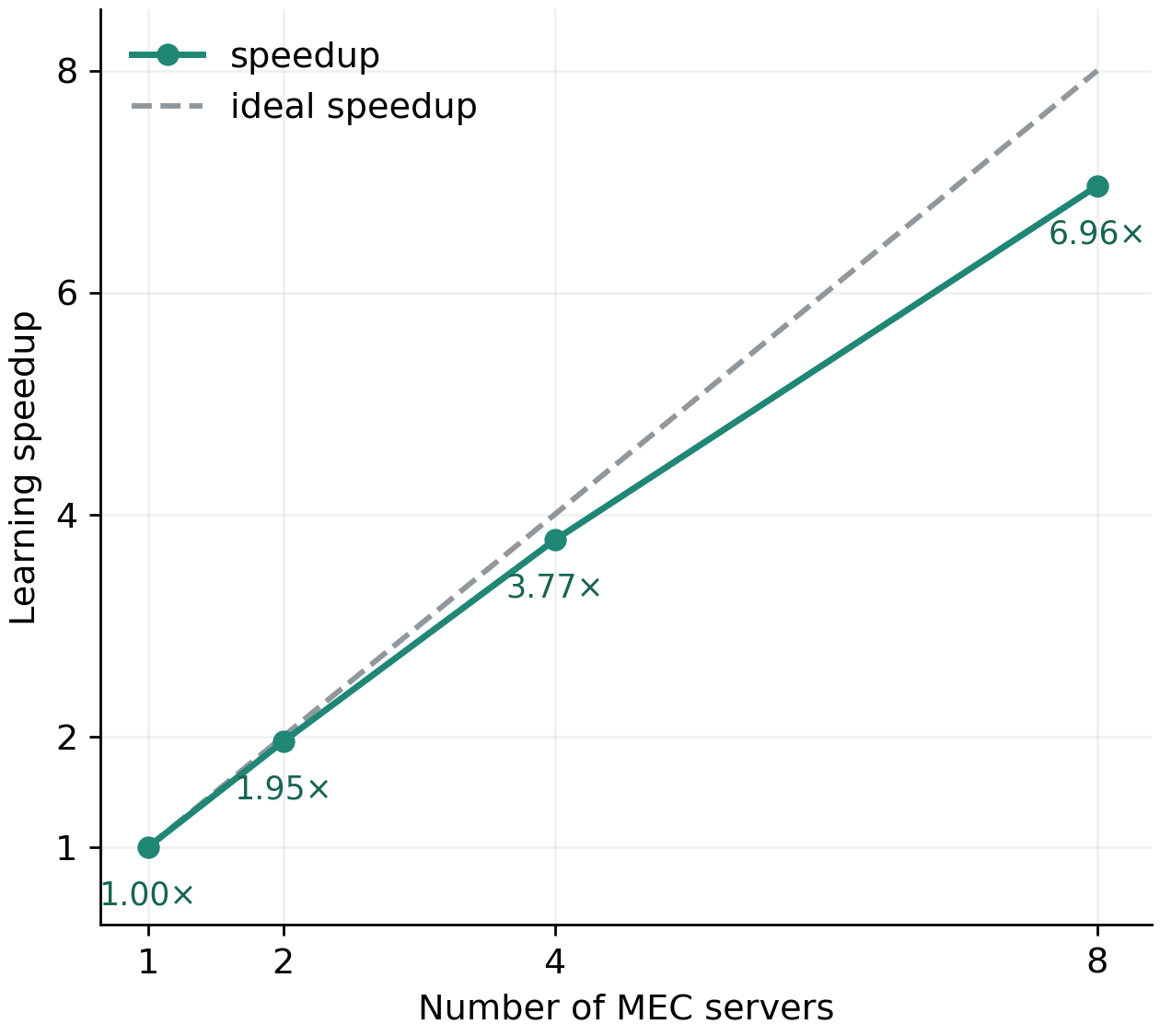}
    \caption{Learning speedup relative to a single independent agent, compared with ideal linear speedup.}
    \label{fig:small_gap_learning_speedup}
\end{figure}

We estimate the learning speedup, which is the ratio of mean number of samples per server in the independent case to the collaborative case \cite{yahya2026service}. Fig.~\ref{fig:small_gap_learning_speedup} compares the measured speedup with the ideal speedup which is proportional to the number of collaborating servers. The observed speedups are 1.95, 3.77, and 6.96 for two, four, and eight CPUs, respectively.

The deviation from ideal linear speedup may result from initialization, server's sampling decisions, and delayed information sharing. Changing the communication threshold $D_{\text{th}}$ controls the local computation and communication tradeoff, which can affect the speedup.

\section{Conclusion}
\label{sec:conclusion}
In this paper, we proposed an approach to reduce the DT-update delays in IoT applications with large computational workloads, where the DT can be updated using a subset of the IoT data  or some information extracted from it. We considered sensing tasks belonging to multiple classes, with different preprocessing demands and freshness requirements. The proposed approach relies on mutual support between the DT and MEC servers: the DT constructs freshness-feasible CPU allocations, while the MEC servers collaboratively apply DistLinGapE to identify an allocation with minimum expected freshness-normalized execution delay. The numerical results show the effectiveness of this approach, while showing that learning complexity depends on the candidate CPU allocations. Collaboration significantly reduces the number of learning rounds, achieving near-linear speedup in the evaluated scenarios. Future work will investigate collaboration among heterogeneous MEC servers, and a better utilization of the CPU resources for further delay reduction.

\bibliography{references}
\bibliographystyle{ieeetr}

\end{document}